\documentclass[12pt,a4paper]{article}
\usepackage{doublespace}
\usepackage{amscd}
\usepackage{amsmath}
\newcommand{\beq}{\begin{equation}}   
\newcommand{\eeq}{\end{equation}}   
\newcommand{\bea}{\begin{eqnarray}}   
\newcommand{\eea}{\end{eqnarray}}   
   
\begin{document}      
\title{Geometric Phases} 
\author{P\'eter L\'evay\\Department of Theoretical Physics, Institute of Physics\\Budapest University of Technology and Economics, Hungary} 
\maketitle

\begin{spacing}{2}
\section{Introduction} 
We invite the reader to perform the following simple experiment.
Put your arm out in front of you keeping your thumb pointing up perpendicular to your arm. Move your arm up over your head, then bring it down to your side, and at last
bring the arm back in front of you again. 
In this experiment an object (your thumb) was taken along a closed path traced by another object (your arm) in a way that a simple local law of transport was applied.
In our case the local law consisted of two ingredients: (1) preserve  the orthogonality of your thumb with respect to your arm and (2) do not rotate the thumb about its instantaneous axis (i.e your arm).  
Performing the experiment, in this way you will manage to avoid rotations of your thumb {\it locally}, however in the end you will experience
a rotation of $90^{\circ}$ {\it globally}.

The experiment above can be regarded as the archetypical example of the phenomenon called {\it anholonomy} by physicists and {\it holonomy} by mathematicians.
In this paper we consider the manifestation of this phenomenon in the realm of quantum theory.
The objects to be transported along closed paths in suitable manifolds will be wave functions representing quantum systems.
After applying {\it local laws} dictated by inputs coming from physics, one ends up with a new wave function that has picked up a complex phase factor.
Phases of this kind are called Geometric Phases 
with the famous Berry Phase being a special case.

\section{The space of rays}

Let us consider a quantum system with physical states represented by elements $\vert\psi\rangle$ of some Hilbert space ${\cal H}$ with scalar product $\langle\vert\rangle:{\cal H}\times{\cal H}\rightarrow {\bf C}$.
For simplicity we assume that ${\cal H}$ is finite dimensional
${\cal H}\simeq {\bf C}^{n+1}$ with $n\geq 1$. The infinite dimensional case can be studied by taking the inductive limit $n\to\infty$.
Let us denote the complex amplitudes characterizing the state $\vert\psi\rangle$
by $Z^{\alpha}, \alpha=0,1,\dots n$. For a normalized state we have

\beq
\label{physics}
\vert\vert\psi\vert\vert^2=\langle\psi\vert\psi\rangle\equiv {\delta}_{\alpha\beta}\overline{Z}^{\alpha}Z^{\beta}\equiv \overline{Z}_{\alpha}Z^{\alpha}=1,
\eeq
\noindent
where summation over repeated indices is understood, indices raised and lowered by ${\delta}^{\alpha\beta}$ and ${\delta}_{\alpha\beta}$ respectively, and the overbar refers to complex conjugation.
A normalized state lies on the unit sphere ${\cal S}\simeq S^{2n+1}$ in ${\bf C}^{n+1}$.
Two nonzero states $\vert\psi\rangle$ and $\vert \varphi\rangle$ are {\it equivalent} $\vert\psi\rangle\sim\vert\varphi\rangle$ iff they are related as $\vert\psi\rangle={\lambda}\vert\varphi\rangle$ for some nonzero complex number ${\lambda}$.
For equivalent states quantities like

\beq
\label{phys}
\frac{\langle\psi\vert {\bf A}\vert \psi\rangle}{\langle\psi\vert\psi\rangle},\quad
\frac{{\vert\langle\psi\vert\varphi\rangle\vert}^2}{{\vert\vert\psi\vert\vert}^2{\vert\vert\varphi\vert\vert}^2},
\eeq

\noindent
having physical meaning (mean value of a physical quantity represented by a Hermitian operator ${\bf A}$, transition probability from a physical state  represented by $\vert\psi\rangle$ to a one represented by $\vert\varphi\rangle$) are invariant. Hence the real space of states representing the physical states of a quantum system unambiguously is the set of equivalence classes ${\cal P}\equiv{\cal H}/\sim$.
${\cal P}$ is called the {\it space of rays}. For ${\cal H}\simeq {\bf C}^{n+1}$
we have ${\cal P}\simeq {\bf CP}^n$, where ${\bf CP}^n$ is the $n$ dimensional complex projective space.
For normalized states $\vert\psi\rangle$ and $\vert\varphi\rangle$ are equivalent iff $\vert\psi\rangle=\lambda\vert\varphi\rangle$, where $\vert\lambda\vert =1$ i.e. $\lambda\in U(1)$.
In words: two normalized states are equivalent iff they differ merely in a complex phase. 
It is well-known that ${\cal S}$ can be regarded as the total space of a principal bundle over ${\cal P}$ with structure group $U(1)$.
This means that we have the projection

\beq
\pi : {\vert\psi\rangle}\in{\cal S}\subset{\cal H}\rightarrow \vert\psi\rangle\langle\psi\vert\in{\cal P},
\eeq
\noindent
where the rank one projector $\vert\psi\rangle\langle\psi\vert$ represents the equivalence class of $\vert\psi\rangle$.
Since we will use this bundle frequently we call it ${\eta}_1$ (the meaning of the subscript $1$ will be clear later). Then we have

\beq
\label{etaone}
{\eta}_1 : U(1)\hookrightarrow {\cal S}\stackrel{\pi}{\longrightarrow} {\cal P}.
\eeq

For $Z^0\neq 0$ our space of rays ${\cal P}$ can be given local coordinates 

\beq
\label{localcoord}
w^j\equiv Z^j/Z^0, \quad j=1,\dots n.
\eeq
\noindent
The $w^j$ are inhomogeneous coordinates for ${\bf CP}^n$ on the coordinate patch ${\cal U}_0$ defined by the condition $Z^0\neq 0$.

${\cal P}$ is a compact complex manifold with a natural Riemannian metric $g$.
This metric $g$ is induced from the scalar product on ${\cal H}$. Let us motivate the construction of $g$ by using the physical input provided by the invariance of the transition probability of (\ref{physics}).
For this we define a {\it distance} between $\vert\psi\rangle\langle\psi\vert$ and $\vert\varphi\rangle\langle\varphi\vert$ in ${\cal P}$ as follows

\beq
\label{distance}
{\cos}^2(\delta(\psi,\varphi)/2)\equiv
\frac{{\vert\langle\psi\vert\varphi\rangle\vert}^2}{{\vert\vert\psi\vert\vert}^2{\vert\vert\varphi\vert\vert}^2}.
\eeq
\noindent
This definition makes sense since
due to the Cauchy-Schwartz inequality the right hand side of (\ref{distance})
is nonnegative and  less than or equal to one. It is equal to one iff $\vert\psi\rangle$ is a nonzero complex multiple of $\vert\varphi\rangle$ i.e. iff they
define the same point in ${\cal P}$. Hence in this case ${\delta}(\psi,\varphi)=0$ as we expected.

Suppose now that $\vert\psi\rangle$ and $\vert\varphi\rangle$ are separated by an infinitesimal distance $ds\equiv \delta(\psi,\varphi)$. Putting this into the definition (\ref{distance}), using the local coordinates $w^j$ of (\ref{localcoord}) for $\vert\psi\rangle$ and $w^j+dw^j$ for $\vert\varphi\rangle$ after Taylor expanding both sides one gets

\beq
\label{FubiniStudy}
ds^2=4g_{j\overline{k}}dw^jdw^{\overline{k}}, \quad j,\overline{k}=1,2,\dots n,
\eeq
\noindent
where
\beq
\label{explicitly}
g_{j\overline{k}}\equiv\frac{(1+\overline{w}_lw^l){\delta}_{jk}-\overline{w}_jw_k}{(1+\overline{w}_mw^m)^2}, 
\eeq
\noindent
with $dw^{\overline{k}}\equiv d\overline{w}^k$.
The line element (\ref{FubiniStudy}) defines the {\it Fubini-Study} metric for ${\cal P}$. 

\section{The Pancharatnam connection}

Having defined our basic entity the space of rays ${\cal P}$ and the principal $U(1)$ bundle ${\eta}_1$ now we define a connection
giving rise to a local law of parallel transport.
This approach gives rise to a very general definition of the geometric phase.
In the mathematical literature the connection we are going to define is called the canonical connection on our principal bundle.
However, since our motivation is coming from physics we are going to rediscover
this construction using merely physical information provided by quantum theory
alone.

The information we need is an adaptation of Pancharatnam's study of polarized light to quantum mechanics.
Let us consider two normalized states $\vert\psi\rangle$ and $\vert\varphi\rangle$.
When these states belong to the same ray, then we have $\vert\psi\rangle =e^{i\phi}\vert\varphi\rangle$ for some phase factor $e^{i\phi}$, hence the phase difference between them can be defined to be just $\phi$.
How to define the phase difference between
$\vert\psi\rangle$ and $\vert\varphi\rangle$ (not orthogonal) when these states belong to {\it different rays}? 
To compare the phases of nonorthogonal states belonging to different rays
Pancharatnam employed the following simple rule:
two states are "in phase" iff their interference is maximal.
In order to find the state $\vert\varphi\rangle\equiv e^{i\phi}\vert\varphi^{\prime}\rangle$ from the ray spanned by the representative $\vert\varphi^{\prime}\rangle$  which is "in phase" with $\vert\psi\rangle$ we have to find a $\phi$ modulo $2\pi$ for which the interference term in

\beq
\label{inter}
{\vert\vert\psi+e^{i\phi}{\varphi}^{\prime}\vert\vert}^2=2(1+{\rm Re}(e^{i\phi}\langle\psi\vert{\varphi}^{\prime}\rangle))
\eeq
\noindent
is maximal.
Obviously the interference is maximal iff
$e^{i\phi}\langle\psi\vert{\varphi}^{\prime}\rangle$ is a real positive number
i.e.

\beq
\label{connection}
e^{i\phi}=\frac{\langle{\varphi}^{\prime}\vert\psi\rangle}{\vert\langle{\varphi}^{\prime}\vert\psi\rangle\vert},\quad \vert\varphi\rangle=\vert{\varphi}^{\prime}\rangle
\frac{\langle{\varphi}^{\prime}\vert\psi\rangle}{\vert\langle{\varphi}^{\prime}\vert\psi\rangle\vert}.
\eeq
\noindent
Hence for the state $\vert\varphi\rangle$ "in phase" with $\vert\psi\rangle$
we have 

\beq
\label{conn}
\langle\psi\vert\varphi\rangle=\vert\langle\psi\vert{\varphi}^{\prime}\rangle\vert\in {\bf R}^{+}.
\eeq

When such $\vert\psi\rangle$ and $\vert\varphi\rangle\equiv\vert\psi +d\psi\rangle$ are infinitesimally separated,
from (\ref{conn}) it follows
that

\beq
\label{inficonnection}
{\rm Im}\langle\psi\vert d\psi\rangle = \frac{1}{2i}\left(\overline{Z}_{\alpha}dZ^{\alpha}-d\overline{Z}_{\alpha}Z^{\alpha}\right)=0
\eeq
\noindent
where $\overline{Z}_{\alpha}Z^{\alpha}=\overline{Z}_0Z^0(1+\overline{w}_jw^j)=1$ due to normalization.
Writing $Z^0\equiv\vert Z^0\vert e^{i\Phi}$ using (\ref{localcoord}) we obtain

\beq
\label{pullback}
{\rm Im}\langle\psi\vert d\psi\rangle=d\Phi+A=0,\quad A\equiv{\rm Im}\frac{\overline{w}_jdw^j}{1+\overline{w}_kw^k}.
\eeq
\noindent
In order to clarify the meaning of the one-form $A$ notice that the choice

\beq
\label{section}
\vert{\psi}^{\prime}\rangle\equiv \frac{1}{\sqrt{1+\overline{w}_kw^k}}
\begin{pmatrix}1\\w^j\end{pmatrix}
\eeq
\noindent
defines a local section of the bundle ${\eta}_1$.
In terms of this section the state $\vert\psi\rangle$ can be expressed as

\beq
\vert\psi\rangle=\begin{pmatrix}Z^0\\Z^j\end{pmatrix}=
\vert Z^0\vert e^{i\Phi}\begin{pmatrix}1\\w^j\end{pmatrix}=e^{i\Phi}\vert{\psi}^{\prime}\rangle.
\eeq
\noindent
For a path $w^j(t)$ lying entirely in ${\cal U}_0\subset{\cal P}$, $\vert\psi(t)\rangle =
e^{i\Phi (t)}\vert{\psi}^{\prime}(t)\rangle$ defines a path in ${\cal S}$
with a $\Phi(t)$ satisfying the equation $\dot{\Phi}+A=0$.
For a closed path $C$ the equation above defines a
 (generically) open path ${\Gamma}$ projecting onto $C$ by the projection
 $\pi$. It must be clear by now that the process we have described is the one of parallel transport with respect to a connection with a connection one-form
 $\omega$. The pull-back of $\omega$ with respect to the (\ref{section}) section
 is the (\ref{pullback}) one-form ($U(1)$ gauge-field) $A$.
 The curve ${\Gamma}$ corresponding to $\vert\psi(t)\rangle$ is the {\it horizontal lift} of $C$ in ${\cal P}$. 
The $U(1)$ phase

\beq
\label{holonomy}
e^{i\Phi[C]}\equiv e^{-i\oint_CA}
\eeq
\noindent
is the holonomy of the connection.
We call this connection the {\it Pancharatnam connection},
and its holonomy for a closed path in the space of rays is the {\it geometric phase} acquired by the wave function.
Now the question of fundamental importance is: how to realize closed paths in ${\cal P}$ physically? This is the question we address in the next sections.

\section{Quantum jumps}

We have seen that physical states of a quantum system are represented by the space of rays ${\cal P}$, and normalized states used as representatives for such states form the total space ${\cal S}$ of a principal $U(1)$ bundle ${\eta}_1$ over
${\cal P}$.
Moreover, in the previous section we have realized that the physical notions
of transition probability, and quantum interference lead us naturally
to the introduction of a Riemannian metric $g$ and an Abelian $U(1)$ gauge-field $A$ living on ${\cal P}$.

An interesting result based on the connection between $g$ and $A$ concerns a nice geometric description of a special type of quantum evolution consisting of a sequence of "quantum jumps".

Consider two nonorthogonal rays $\vert A\rangle\langle A\vert$ and $\vert B\rangle\langle B\vert$ in ${\cal P}$. 
Let us suppose that the system's normalized wave function initially is $\vert A\rangle\in {\cal S}$, and measure it by the "polarizer" $\vert B\rangle\langle B\vert$. Then the result of this filtering measurement is $\vert B\rangle\langle B\vert A\rangle$, or after projecting back to the set of normalized states
we have the "quantum jump"

\beq
\label{jump}
\vert A\rangle \rightarrow \vert B\rangle\frac{\langle B\vert A\rangle}{\vert\langle B\vert A\rangle\vert}.
\eeq
\noindent
Now we have the following: 

{\it Theorem}: The (\ref{jump}) jump can be recovered by parallel transporting the normalized state $\vert A\rangle$ according to the Pancharatnam connection along the {\it shortest geodesic} (with respect to the (\ref{explicitly}) metric), connecting $\vert A\rangle\langle A\vert$ and $\vert B\rangle\langle B\vert$ in ${\cal P}$. 

Let us now consider a cyclic series of filtering measurements with projectors
$\vert A_a\rangle\langle A_a\vert$ , $a=1,2,\dots N+1$, where $\vert A_1\rangle\langle A_1\vert=\vert A_{N+1}\rangle\langle A_{N+1}\vert$.
Prepare the system in the state $\vert A_1\rangle\in {\cal S}$, and then subject it to the sequence of filtering measurements. Then according to the theorem, the phase

\beq
\label{polygon}
e^{i{\Phi}}=\frac{\langle A_1\vert A_N\rangle\langle A_N\vert A_{N-1}\rangle\cdots\langle A_2\vert A_1\rangle}
{\vert\langle A_1\vert A_N\rangle\langle A_N\vert A_{N-1}\rangle\cdots\langle A_2\vert A_1\rangle\vert}
\eeq
\noindent
picked up by the state equals to the one obtained by parallel transporting $\vert A_1\rangle$ along a geodesic polygon  consisting of the shorter arcs
connecting the projectors $\vert A_a\rangle\langle A_a\vert$ and $\vert A_{a+1}\rangle\langle A_{a+1}\vert$ with $a=1,2,\dots N$.
It is important to realize that this filtering measurement process is {\it not a unitary one}, hence unitarity is not essential for the geometric phase to appear.

In this subsection we have managed to obtain closed paths in the form of {\it geodesic polygons} in ${\cal P}$ via the physical process of subjecting the initial state $\vert A_1\rangle$ to a sequence of filtering measurements.
It is clear that for any type of evolution the geodesics of the Fubini-Study metric play a fundamental role since any smooth closed curve in ${\cal P}$ can be approximated by geodesic polygons. 

Nonunitary evolution provided by the quantum measuring process
is only half of the story.
In  the next section we start describing closed paths in ${\cal P}$
arising also from unitary evolutions generated by parameter dependent Hamiltonians, the original context where geometric phases were discovered.

\section{Unitary evolutions}

\subsection{Adiabatic evolution}
Suppose that the evolution of our quantum system with ${\cal H}\simeq {\bf C}^{n+1}$ is generated by a Hermitian Hamiltonian matrix depending on a set of external parameters $x^{\mu}, \mu = 1,2\dots M$.
Here we assume that the $x^{\mu}$ are local coordinates on some coordinate patch ${\cal V}$ of a smooth $M$ dimensional manifold ${\cal M}$.
We label the eigenvalues of $H(x)$ by the numbers $r=0,1,2,\dots n$, and assume that the $r$ th eigenvalue $E_r(x)$ is nondegenerate. 
We have

\beq
\label{eigen}
H(x)\vert r,x\rangle= E_r(x)\vert r, x\rangle,\quad r=0,1,2,\dots n.
\eeq
\noindent
We assume that $H(x), E_r(x), \vert r, x\rangle$ are smooth functions of $x$.
The rank one spectral projectors

\beq
\label{projectors}
P_r(x)\equiv \vert r, x\rangle\langle r, x \vert. \quad r=0,1,2,\dots n
\eeq
\noindent
for each $r$ define a map $f_r: {\cal M}\rightarrow {\cal P}$

\beq
\label{map}
f_r: x\in {\cal V}\subset {\cal M}\mapsto P_r(x)\in{\cal P}.
\eeq
\noindent

Recall now that we have the bundle ${\eta}_1$ over ${\cal P}$ at our disposal,
and we can pull back ${\eta}_1$ using the map $f_r$ to construct a new bundle ${\xi}_1^r$ over the parameter space ${\cal M}$. Moreover, we can define a connection on ${\xi}_1^r$ by pulling back the canonical (Pancharatnam) connection  of ${\eta}_1$.
The resulting bundle ${\xi}_1^r$ is called the Berry-Simon bundle over the parameter space ${\cal M}$.
Explicitly we have the bundle

\beq
\label{BS}
{\xi}_1^r : U(1)\hookrightarrow {\xi}_1^r\stackrel{{\pi}_{\xi}}{\longrightarrow} {\cal M}.
\eeq
\noindent
The states $\vert r, x\rangle$ of
(\ref{eigen}) define a local section of ${\xi}_1^r$.
Supressing the index $r$ the relationship between ${\eta}_1$ and ${\xi}_1$ can be summarized by the following diagram

\beq
\label{diagram}
\begin{CD}
{{\xi}_1}@<{f^{\ast}}<<{{\eta}_1}\\
@V{{\pi}_{\xi}}VV@V{{\pi}_{\eta}}VV\\
{\cal M}@>f>>{\cal P}
\end{CD}
\eeq
\noindent
Here $f^{\ast}$ denotes the pullback map, and we have ${\xi}_1 \equiv f^{\ast}({\eta}_1)$. (We have denoted the total space ${\cal S}$ as ${\eta}_1$.)

The local section of ${\xi}_1$ arising as the pull back of the (\ref{section})
one of ${\eta}_1$ is given by

\beq
\label{pullsection}
\vert r,x\rangle =\frac{1}{\sqrt{1+\overline{w}^k(x)w_k(x)}}\begin{pmatrix}1\\w^j(x)\end{pmatrix},\quad x\in{\cal V}\subset {\cal M},
\eeq
\noindent
with $j=1,2,\dots n$.
The pullback of the Pancharatnam connection $\omega$ on ${\eta}_1$ is $f^{\ast}({\omega})$. We can further pull back 
$f^{\ast}({\omega})$ to ${\cal V}\subset{\cal M}$ with respect to the (\ref{pullsection}) section to obtain a gauge field living on the parameter space.
This gauge-field is called the {\it Berry gauge field} and the corresponding connection is the {\it Berry connection}. We have

\beq
\label{Berry}
{\cal A}=f^{\ast}(A)={\cal A}_{\mu}(x)dx^{\mu}=(A_j{\partial}_{\mu}w^j+A_{\overline{j}}{\partial}_{\mu}w^{\overline{j}})dx^{\mu}.
\eeq
\noindent
Here ${\partial}_{\mu}\equiv\frac{\partial}{{\partial}x^{\mu}}$ and $A$ is given by
(\ref{pullback}). 
When we have a closed curve ${\cal C}$ in ${\cal M}$ then  $f\circ {\cal C}$ defines a closed curve $C$ 
in ${\cal P}$. We already know that the holonomy for $C$ in ${\cal P}$ can be written in the (\ref{holonomy}) form hence we can write

\beq
\label{Phase}
{\Phi}_B=-\oint_{f\circ {\cal C}}A=-\oint_{\cal C}f^{\ast}(A)=-\oint_{\cal C}{\cal A}.
\eeq
\noindent
This formula says that there is a  geometric phase picked up by the eigenstates
of a parameter dependent Hermitian Hamiltonian when we
change the parameters along a closed curve.
Our formula shows that the geometric phase can be either calculated using the canonical connection on ${\eta}_1$ or the Berry connection on ${\xi}_1$.

Let us then change the parameters $x^{\mu}$ adiabatically. The closed path in parameter space then defines Hamiltonians satisfying $H(x(T))=H(x(0))$ for some $T\in {\bf R}^{+}$. Moreover, there is also the associated closed curve $P_r(x(T))=P_r(x(0))$ in ${\cal P}$. 
The quantum adiabatic theorem
states that if we prepare a state $\vert\Psi(0)\rangle\equiv\vert r, x(0)\rangle$ at $t=0$ which is an eigenstate of the instantaneous Hamiltonian $H(x(0))$, then under changing the parameters infinitely slowly the time evolution generated by the time dependent Schr\"odinger equation

\beq
\label{Sch}
i\hbar\frac{d}{dt}\vert\Psi(t)\rangle=H(t)\vert\Psi(t)\rangle
\eeq
\noindent
takes  this  after time $t$ into

\beq
\label{Fock}
\vert\Psi(t)\rangle=\vert r, x(t)\rangle e^{i{\Lambda}_r(t)}
\eeq
\noindent
which belongs to the {\it same} eigensubspace.
The point is that the theorem holds only for cases when the kinetic energy associated with the slow change in the external parameters is much smaller then the energy separation between $E_r(x)$ and $E_{r^{\prime}}(x)$ for all $x\in {\cal M}$. Under this assumption during the evolution transitions between adjacent levels are prohibited. Notice that the adiabatic theorem clearly breaks down in the vicinity of level crossings where the gap is comparable with the magnitude of the kinetic energy of the external parameters. 

However, 
if one takes it for granted that the projector $P_r(t)\equiv P_r(x(t))$ for some 
$r$ satisfies the Schr\"odinger-von Neumann equation 

\beq
i\hbar\frac{d}{dt}P_r(t)=[H(t), P_r(t)],
\eeq
\noindent
by virtue of (\ref{eigen}) we get zero for the right hand side. This means that $P_r(t)$ is constant, 
hence our curve in ${\cal P}$ degenerates to a point.
The upshot of this is that exact adiabatic cyclic evolutions do not exist.
It can be shown however, that under certain conditions one can find an initial state $\vert\Psi(0)\rangle\neq \vert r, x(0)\rangle$ that is "close enough"
to $P_r(x(t))=\vert r, x(t)\rangle\langle r, x(t)\vert$.
Then we can say that the projector analogue of (\ref{Fock}) only approximately holds

\beq
\vert\Psi(t)\rangle\langle \Psi(t)\vert \simeq \vert r, x(t)\rangle\langle r, x(t)\vert.
\eeq
\noindent
This means that our use of the bundle picture for the generation of closed curves for ${\cal P}$ via the adiabatic evolution can merely be used as an approximation.

\subsection{Berry's Phase}

Let us substitute (\ref{Fock}) into (\ref{Sch}). Then straightforward calculation shows that

\beq
\label{BP}
e^{i{\Lambda}_r(T)}=e^{-\frac{i}{\hbar}\int_0^TE_r(t)dt}e^{-i\oint_{\cal C}{\cal A}^{(r)}}
\eeq
\noindent
where ${\cal C}$ is a closed curve lying entirely in ${\cal V}\subset {\cal M}$.
The first phase factor is the {\it dynamical} and the second is the celebrated {\it Berry Phase}. 
Notice that the index $r$ labeling the eigensubspace in question should now be included in the (\ref{Berry}) definition of ${\cal A}$ .

As an explicit example let us take the Hamiltonian

\beq
\label{spin}
H({\bf X}(t))=-{\omega}_0 {\bf J}{\bf X}(t),\quad {\omega}_0\equiv\frac{Bge}{2mc},\quad \bf X\in {\bf R}^3,\quad \vert{\bf X}\vert ={\rm 1}
\eeq
\noindent
where $e$, $m$ and $g$ are the charge, mass and Land\'e factor of a particle, $c$ is the speed of light and $B$ is the (constant) magnitude of an applied magnetic field. The three components of ${\bf J}$ are $(2J+1)\times(2J+1)$ dimensional spin matrices satisfying ${\bf J}\times{\bf J}=i\hbar{\bf J}$.
The (\ref{spin}) Hamiltonian describes a spin $J$ particle moving in a magnetic field with slowly varying direction. 
It is obvious that the parameter space is a two-sphere. Introducing polar coordinates $0\leq\theta<\pi$, $0\leq\chi<2\pi$ for the patch ${\cal V}$ of $S^2$ excluding the south pole,
we have $x^1\equiv\theta$, $x^2\equiv\chi$.

As an illustration let us consider the spin $\frac{1}{2}$ case. Then $H$ can be expressed in terms of the $2\times 2$ Pauli matrices. The eigenvalues are $E_0=-{\omega}_0\hbar/2$ and $E_1={\omega}_0\hbar/2$, ($r=0,1$).
For the ground state the mapping $f_{0}$ of (\ref{map}) from ${\cal V}\subset{\cal M}\simeq S^2$ to ${\cal P}\simeq {\bf CP}^1$ is  given by

\beq
\label{expl}
w(\theta, \chi)\equiv \tan\left(\frac{\theta}{2}\right)e^{i\chi}
\eeq
\noindent
which is stereographic projection of $S^2$ from the south pole onto the complex plane corresponding to the coordinate patch ${\cal U}_0\subset {\bf CP}^1$.
Using (\ref{pullback}) and (\ref{Berry}) one can calculate the pullback gauge field and its curvature ${\cal F}^{(0)}\equiv d{\cal A}^{(0)}$

\beq
\label{result2}
{\cal A}^{(0)}=\frac{1}{2}(1-\cos\theta)d\chi,\quad {\cal F}^{(0)}=\frac{1}{2}\sin\theta d\theta\wedge d\chi.
\eeq
\noindent
Notice that ${\cal F}^{(0)}$ is the field strength of a magnetic monopole of strength $\frac{1}{2}$ living on ${\cal M}$. 
Using Stokes theorem from (\ref{Phase}) one can calculate Berry's Phase

\beq
\label{BerryPh}
{\Phi}^{(0)}[{\cal C}]=-\oint_{\cal C}{\cal A}^{(0)}=-\int_S{\cal F}^{(0)}=-\frac{1}{2}{\Omega}[{\cal C}],
\eeq
\noindent
where $S$ is the surface bounded by the loop ${\cal C}$ and  ${\Omega}[{\cal C}]$ is the solid angle the curve ${\cal C}$ subtends at $\bf X={\bf 0}$.

The above result can be generalized for arbitrary spin $J$.
Then we have the eigenvalues $E_r=-{\omega}_0\hbar (J-r)$ where $0\leq r\leq 2J$.
The final result in this case is

\beq
\label{general}
{\Phi}^{(r)}[{\cal C}]=-(J-r){\Omega}[{\cal C}], \quad 0\leq r\leq 2J.
\eeq
\noindent

\subsection{The Aharonov-Anandan Phase}

We have seen that the quantum adiabatic theorem can only be used approximately for generating closed curves in ${\cal P}$. 
In this section we
show how to generate such curves exactly.

Let us consider the (\ref{Sch}) Schr\"odinger equation with a time dependent Hamiltonian.
Then we call its solution $\vert\Psi(t)\rangle$ {\it cyclic} if the state of the system returns after a period $T$ to its original state. This means that the projector $\vert\Psi(t)\rangle\langle \Psi(t)\vert$ traverses a closed path $C$ in ${\cal P}$. In order to realize this situation we have to find solutions of (\ref{Sch}) for which $\vert\Psi(T)\rangle=e^{i{\Delta}_{\Psi}}\vert\Psi(0)\rangle$ for some ${\Delta}_{\Psi}$.

Taking for granted the existence of such a solution, let us first explore its consequences. First we remove the dynamical phase from the cyclic solution $\vert\Psi(t)\rangle$ 

\beq
\label{remove}
\vert \psi(t)\rangle \equiv e^{\frac{i}{\hbar}\int_0^t\langle \Psi(t^{\prime})\vert H(t)\vert\Psi(t^{\prime})\rangle dt^{\prime}}\vert\Psi (t)\rangle.
\eeq
\noindent
Then $\vert\psi(t)\rangle$ satisfies (\ref{inficonnection}) i.e. it defines a {\it unique} horizontal lift of the closed curve $C$ in ${\cal P}$. Following the same steps as in Section III. we see that the phase

\beq
\label{AA} 
{\Phi}_{AA}[C]=-\oint_CA={\Delta}_{\Psi}+\frac{1}{\hbar}\int_0^T
\langle \Psi(t)\vert H(t)\vert\Psi(t)\rangle dt
\eeq
\noindent
is of purely geometric in origin. It is called the {\it Aharonov-Anandan} (AA).
{\it Phase}.

Let us now turn back to the question of finding cyclic states satisfying
$\vert\Psi(T)\rangle =e^{i\Delta_{\Psi}}\vert\Psi(0)\rangle$.
One possible solution is as follows.
Suppose that $H$ depends on time through some not necessarily slowly changing parameters $x$.
Let us find a partner Hamiltonian $h$ for our $H$ by defining a smooth mapping $\sigma:{\cal M}\rightarrow {\cal M}$, such that
\beq
\label{smooth}
h(x)\equiv H(\sigma(x)), \quad x\in {\cal V}\subset{\cal M}.
\eeq
\noindent
For the special class we will study the {\it cyclic vectors are eigenvectors of}
$h(x)$.
Hence the projectors $p_r$ and $P_r$ of $h$ and $H$ are related as
$p_r(x)=P_r(\sigma(x))$ this means that we have a map $g_r: {\cal M}\rightarrow {\cal P}$

\beq
\label{g}
g_r\equiv f_r\circ \sigma: x\in {\cal V}\subset {\cal M}\rightarrow p_r(x)\in{\cal P}
\eeq
\noindent
which associates to every $x$ an eigenstate of $h(x)$. Moreover $g_r$ associates to a closed curve ${\cal C}$ in ${\cal M}$ a closed curve $C$ in ${\cal P}$.
Notice that generically $[h(x),H(x)]\neq 0$ hence cyclic states
are not eigenstates of the instanteneous Hamiltonian.

It should be clear by now that we can repeat the construction as discussed in the adiabatic case with $g_r$ replacing $f_r$. In particular we can construct a new bundle ${\zeta}_1$ over the parameter space via the usual pullback procedure. More precisely we have the corresponding diagram
\beq
\label{diagram2}
\begin{CD}
{{\zeta}_1}@<{g^{\ast}}<<{{\eta}_1}\\
@V{{\pi}_{\zeta}}VV@V{{\pi}_{\eta}}VV\\
{\cal M}@>g>>{\cal P}
\end{CD}
\eeq
\noindent
We can pullback the Pancharatnam connection, yielding the AA connection.

\beq
\label{AAconnection}
a\equiv g^{\ast}(A)={\sigma}^{\ast}\circ f^{\ast}(A)={\sigma}^{\ast}({\cal A}).
\eeq
\noindent
Where the last equality relates the AA connection with the Berry connection.
Now the AA phase is

\beq
\label{Phase2}
{\Phi}_{AA}=-\oint_{g\circ {\cal C}}A=-\oint_{\cal C}g^{\ast}(A)=-\oint_{\cal C} a.
\eeq
\noindent

As an example let us take the (\ref{spin}) Hamiltonian with the curve ${\cal C}$
on ${\cal M}\equiv S^2$ 

\beq
\label{curve}
{\bf X}(t)=(\sin\theta\cos(\chi +\omega t),\sin\theta\sin(\chi +\omega t), \cos\theta).
\eeq
\noindent
Here $\theta$ and $\chi$ are the polar coordinates of a {\it fixed point} in $S^2$ where the motion starts. The curve ${\cal C}$ is a circle of fixed
latitude and is traversed with an arbitrary speed. This model can be solved exactly and it can be shown that the mapping ${\sigma}_s :S^2\rightarrow S^2$ is
given by
\beq
\label{mapping}
\sigma: (u,\chi)\mapsto (\frac{u-s}{\sqrt{s^2-2us +1}},\chi),\quad u\equiv\cos\theta, \quad s\equiv\frac{\omega}{{\omega}_0}.
\eeq
\noindent
One can prove that for $0\leq s<1$ ${\sigma}_s$ is a diffeomorphism. In the $s\to 0$ (the adiabatic) limit the mapping $g_{r,s}\equiv f_{r,s}\circ{\sigma}_s$ is continuously deformed to $f_r$.  Moreover, $h(x)$ as defined above commutes with the time evolution operator hence cyclic states are indeed eigenstates
of $h(x)$.

Using (\ref{AAconnection}), (\ref{Phase2}) and the (\ref{mapping}) explicit form of ${\sigma}_s$ we get for the AA Phase the result

\beq
\label{AAP2}
{\Phi}^{(r,s)}_{AA}[{\cal C}]=-2\pi(J-r)\left(1-\frac{u-s}{\sqrt{s^2-2us+1}}\right)
\eeq
\noindent
In the adiabatic limit the result goes to $-2\pi(J-r)(1-u)$ which is just $-(J-r)$ times the solid angle of our path of fixed latitude, as it has to be.

\section{Generalization}

In our sequence of examples we have shown that geometric phases are related to the geometric structures on  the bundle ${\eta}_1$.
The Berry and AA phases are special cases arising from
Pancharatnam's Phase via a pullback procedure with respect to suitable maps defined by the physical situation in question.
Hence the Pancharatnam connection in this sense is {\it universal}.
The root of this universality rests in a deep theorem of mathematics concerning the existence of {\it universal bundles} and their {\it universal connections}. 
In order to elaborate the insight provided by this theorem into the geometry of quantum evolution let us first make a further generalization.

In our study of time dependent Hamiltonians we have assumed
that the eigenvalues of (\ref{eigen}) were {\it nondegenerate}.
Let us now relax this assumption. Fix an integer $N\geq 1$, 
the degeneracy of the eigensubspace corresponding to the eigenvalue $E_r$.
One can then form a $U(N)$ principal bundle ${\xi}_N$ over ${\cal M}$,
furnished with a connection, that is a natural generalization of the Berry  connection.
The pullback of this connection to a patch of ${\cal M}$ is a $U(N)$-valued gauge-field and its  holonomy along a loop in ${\cal M}$ gives rise to a $U(N)$ matrix generalization of the $U(1)$ Berry phase.  

The natural description of this connection and its AA analogue is as follows.
Take the complex Grassmannian $Gr(n+1, N)$ of $N$ planes in ${\bf C}^{n+1}$.
Obviously $Gr(n+1,1)\equiv {\cal P}$. Each point of $Gr(n+1, N)$ corresponds to an $N$ plane through the origin represented by a rank-$N$ projector.
This projector can be written in terms of $N$ orthonormal basis vectors in an infinity of ways. This ambiguity of chosing orthonormal frames is captured by the $U(N)$ gauge symmetry, the analogue of the $U(1)$ (phase) ambiguity in defining a normalized state as the representative of the rank one projector.  
This bundle of frames is the Stiefel bundle $V(n+1, N)$ alternatively denoted by ${\eta}_N$. $V(n+1, N)$ is a principal $U(N)$ bundle over $Gr(n+1, N)$ equipped with a canonical connection ${\omega}_N$ which is the $U(N)$ analogue of Pancharatnam's connection.

Now according to the powerful theorem of Narasimhan and Ramanan
if we have a $U(N)$ bundle ${\xi}_N$ over the $M$ dimensional parameter space ${\cal M}$ then there exists an integer $n_0(N,M)$ such that for $n\leq n_0$
there exists a map $f: {\cal M}\rightarrow Gr(n+1,N)$ such that ${\eta}_N=f^{\ast}(V(n+1,N))$. Moreover, given any two such maps $f$ and $g$ the corresponding pullback bundles are isomorphic if and only if $f$ is homotopic to $g$.

For the examples of Sections 5.2 and 5.3 we have $N=1$ , $n=1$ and $M=2$ and since the maps $f_{r}$ and $g_{r,s}$  defined by the rank one spectral projectors of $H(x)$ and $h(x)$ for $0\leq s<1$ are homotopic, the corresponding pullback bundles ${\xi}_1$ and ${\zeta}_1$ are isomorphic.  
Moreover, the Berry and AA connections are the pullbacks of the universal connection on $V(n+1,1)\equiv {\eta}_1$ which is just Pancharatnam's connection.

For the infinite dimensional case one can define $Gr(\infty, N)$ by taking the union of the natural inclusion maps of $Gr(n,N)$ into $Gr(n+1, N)$. 
We denote this universal classifying bundle $V(\infty, N)$ as ${\eta}$.
Then we see that given an $N$ dimensional eigensubspace bundle over ${\cal M}$ and a map $f_r: x\in{\cal M}\mapsto P_r(x)\in Gr(\infty, N)$ defined by the physical situation, the geometry of evolving eigensubspaces  can be understood in terms of the holonomy of the pullback of the universal
connection on ${\eta}$.

\section{Conclusions}

In this paper we illuminated the mathematical origin of geometric phases.
We have seen that the key observation is the fact that the space of rays ${\cal P}$ represents unambiguously the physical states of a quantum system.
The particular representatives of a class in ${\cal P}$ belonging to the usual Hilbert space ${\cal H}$ form (local) sections of a $U(1)$ bundle ${\eta}_1$.
Based on the physical notions of transition probability and interference   
${\eta}_1$ can be furnished with extra structures: the metric and the connection, the latter giving rise to a natural definition of parallel transport.
We have seen that the geodesics of ${\cal P}$ with respect to our metric play a fundamental role in approximating evolutions of any kind giving rise to a curve in ${\cal P}$,

The geometric structures of ${\eta}_1$ induce similar structures for pull-back bundles. These bundles encapsulate the geometric details of time evolutions generated by Hamiltonians depending on a set of parameters $x$ belonging to a manifold ${\cal M}$. It was shown that the famous examples of Berry and Aharonov-Anandan Phases arise as an important special case in this formalism.
A generalization of evolving $N$ dimensional subspaces based on the theory of universal connections can also be given.
This shows that the basic structure responsible for the occurrence of anholonomy effects in evolving quantum systems
is the universal bundle ${\eta}$ which is the bundle of subspaces of arbitrary dimension $N$ in a Hilbert space.

We have not touched the important issue of applying the idea of anholonomy
to physical problems. Let us mention here that there are spectacular applications like holonomic quantum computation, the gauge kinematics of deformable bodies,
quantum Hall-effect, fractional spin and statistics etc.
The interested reader should consult the vast literature on the subject or as a first glance the book of A. Shapere and F. Wilczek
listed at the references.

\end{spacing}
\end{document}